\documentclass[fleqn,usenatbib]{mnras}
\usepackage{newtxtext,newtxmath}
\usepackage[T1]{fontenc}
\usepackage{ae,aecompl}
\usepackage{graphicx}	
\usepackage{amsmath}	
\title[Identifying the Blazar Subclasses with XGBoost]{Classification of the \emph{Fermi}-LAT Blazar Candidates of Uncertain type using 
eXtreme Gradient Boosting}

\author[Tolamatti, Singh and Yadav]
{A. Tolamatti,$^{1,2}$\thanks{E-mail: anilkumart@barc.gov.in (AT)}
K. K. Singh,$^{1,2}$\thanks{E-mail: kksastro@barc.gov.in (KKS)}
K. K. Yadav,$^{1,2}$
\\
$^{1}$Astrophysical Sciences Division, Bhabha Atomic Research Centre, Mumbai, 400085, India\\
$^{2}$Homi Bhabha National Institute, Anushakti Nagar, Mumbai, 400094, India
}

\date{Accepted XXX. Received YYY; in original form ZZZ}

\pubyear{2023}

\begin{document}
\label{firstpage}
\pagerange{\pageref{firstpage}--\pageref{lastpage}}
\maketitle
\begin{abstract}
Machine learning based approaches are emerging as very powerful tools for many applications  
including source classification in astrophysics research due to the availability of huge 
high quality data from different surveys in observational astronomy. The Large Area Telescope 
on board \emph{Fermi} satellite (\emph{Fermi}-LAT) has discovered more than 6500 high energy 
gamma-ray sources in the sky from its survey over a decade. A significant fraction of 
sources observed by the \emph{Fermi}-LAT either remains unassociated or has been identified as 
\emph{Blazar Candidates of Uncertain type} (BCUs). We explore the potential of eXtreme Gradient 
Boosting (XGBoost)- a supervised machine learning algorithm to identify the blazar subclasses among 
a sample of 112 BCUs of the 4FGL catalog whose X-ray counterparts are available within 
95$\%$ uncertainty regions of the \emph{Fermi}-LAT observations. We have used information from 
the multi-wavelength observations in IR, optical, UV, X-ray and $\gamma$-ray wavebands along with 
the redshift measurements reported in the literature for classification. Among the 112 uncertain 
type blazars, 62 are classified as BL Lacertae objects (BL Lacs) and 6 have been classified 
as Flat Spectrum Radio Quasars (FSRQs). This indicates a significant improvement with respect to the 
multi-perceptron neural network based classification reported in the literature. 
Our study suggests that the gamma-ray spectral index, and IR color indices are the most important features 
for identifying the blazar subclasses using the \emph{XGBoost} classifier. 
We also explore the importance of redshift in the classification BCU candidates. 
\end{abstract}



\section{Introduction}
Blazars represent a dominant population of high energy $\gamma$-ray sources in the extragalactic Universe. 
They are observed to emit non-thermal radiation over the whole electromagnetic spectrum from 
radio to very high energy gamma rays. Their broadband emission exhibits various observational 
features like frequent occurrence of flaring episodes with strong flux variability at different 
timescales, high degree of polarization and swing in polarization angle, superluminal motion 
of jets, compact radio emission, relativistic beaming and Doppler boosting of radiation. These 
observational properties are very well explained by a standard paradigm which suggests that 
blazars are radio loud active galactic nuclei and their broadband emission originates from a 
relativistic plasma jet oriented at a very small acute angle to the line of sight of the observer 
on Earth \citep{Urry1995,Netzer2015,Dermer2015}. A supermassive black hole, located at the center of 
host galaxy with elliptical morphology, is assumed to launch a pair of highly collimated plasma 
jets extending to large distances (at kiloparsec scale) in opposite directions. These outflows arise 
from the proximity of black hole and are powered by the accretion of low viscosity gaseous matter in 
the Keplerian orbit from the surrounding onto the black hole. The angular momentum of infalling matter, 
being transferred outward, provides a steady and slow inflow to the jet \citep{Blandford1982,Blandford2019}.
\par
In the inner jet regions, particles are accelerated to ultra-high energies and the plasma moves at relativistic 
speed. Therefore, the non-thermal radiation produced in the jet is blue-shifted and Doppler boosting is invoked to 
explain some of the observed features of blazars. A complete understanding of the physical processes involved in the 
non-thermal emission from blazar jets, covering about 15 orders of magnitude in energy, still remains an active area 
of research. This continuum emission is described by a broadband spectral energy distribution (SED) having two 
characteristic humps peaking at low and high energies. The peak of first/low-energy hump lies in the optical, 
ultraviolet or X-ray range and the second/high-energy hump peaks at gamma-ray energies. Thermal emissions from 
the host galaxy, accretion disk and other components also dominate at low energies in the SED of a 
few blazars \citep{Singh2019a}. A range of physically acceptable models are proposed to explain and best fit 
the multi-wavelength data points from different observations of the blazars \citep{Sol2022,Singh2020}. 
Depending on the structure of emission region and radiating particles, the blazar emission models are described as 
single-zone leptonic, single-zone hadronic and multi-zone leptonic/hadronic. The most likely radiative process for 
low energy hump is the widely accepted synchrotron radiation of utrarelativistic electrons in the jet for all 
the models \citep{Giommi2021,Singh2022a}. The origin of high energy peak is still not clear and remains an open question 
in the field of blazar research. In the single zone leptonic models, inverse Compton scattering of synchrotron photons itself 
or external photon field by the relativistic electrons in the emission region is considered as the dominant process for 
producing energetic $\gamma$-ray photons in the blazar jets \citep{Marscher1985,SinghKP2022,Tolamatti2022}. In some cases, 
synchrotron radiation of relativistic protons and photo-hadronic interactions are invoked to reproduce the observed high 
energy hump in the SED of blazars \citep{Aharonian2002,Bottcher2013}. One zone models are often not found to be suitable for 
explaining the broadband SED of blazars and therefore two zone or multi-zone emission models have also been proposed as viable 
alternative \citep{Yamada2020,Sahu2021,Ghosal2022}. In such models, the accelerated particles more or less continuously radiate 
non-thermal emission along the jet and the energy densities associated with radiating particles and jet magnetic field follow a 
power law as a function of distance from central region \citep{Potter2013}.  
\par
Blazars are phenomenologically classified on the basis of observational evidences and key features to develop a better 
understanding of the physical processes involved in the broadband emission. According to the emission/absorption line features 
over the continuum non-thermal optical emission and equivalent width of emission lines in the source rest frame, blazars 
are divided into two classes namely BL Lacaerte objects (BL Lacs) and Flat Spectrum Radio Quasars (FSRQs). Of the two subclasses,  
BL Lacs have either weak or no emission lines whereas FSRQs exhibit strong and broad emission lines \citep{Stocke1991}. 
The accretion disk in FSRQs are observed to be radiatively more efficient than that in BL Lacs and therefore FSRQs have brighter 
accretion disks. The luminosity of broad emission line can also be used as an important feature for blazar 
classification \citep{Ghisellini2011}. Alternatively, blazars are also classified on the basis of the position of peak-frequency 
of the low-energy/synchrotron hump in their characteristic broadband SED. According to this classification, blazars are grouped 
as low-synchrotron peaked (LSP), intermediate-synchrotron peaked (ISP) and high-synchrotron peaked (HSP) sources \citep{Abdo2010}. 
Most of the FSRQs are LSP with their synchrotron peak frequency below $10^{14}$ Hz whereas synchrotron peak is positioned between $10^{14}$ Hz and 
$10^{17}$ Hz for BL Lacs which are mostly HSP. LSP, ISP and HSP BL Lac sources are also referred to as LBL, IBL and HBL respectively in literature. 
A small fraction of blazars exhibit synchrotron peak frequency above $10^{17}$ Hz and are referred extremely high-synchrotron peaked (EHSP) sources or EHBL 
\citep{Costamante2001,Singh2019b}.  The high energy observations in the energy range above 100 MeV suggest that FSRQs are characterized, 
on an average, by softer $\gamma$-ray spectra than BL Lacs \citep{Ghisellini2009}. Consequently, BL Lacs are considered to be the most extreme 
TeV gamma-ray sources. The cosmological evolution of blazars suggests that the distribution of the population of FSRQs peaks at higher redshift than 
that of the BL Lacs observed so far \citep{Ackermann2015}. This implies that BL Lacs are relatively younger objects than FSRQs in the extragalactic Universe. 
\par
Dedicated multi-wavelength monitoring programs involving different instruments world-wide over the last two decades have greatly helped in 
better understanding of the blazars. However, the synergy and link between blazar subtypes remain poorly understood even today. Launch of 
Large Area Telescope (LAT) onboard the \emph{Fermi} satellite in 2008 \citep{Atwood2009,Ajello2021} has revolutionized the field of observational 
high energy astrophysics with the successive release of several point source catalogs containing thousands of gamma-ray sources 
\citep{Abdo2010,Ackermann2016,Ajello2017,Abdollahi2020,von2020}. Majority of the high altitude sources detected by the \emph{Fermi}-LAT in the energy 
range from 100 MeV to close to 1 TeV are blazars. With the detection of gamma-ray emission from more than 3000 blazars at high confidence level, 
the \emph{Fermi}-LAT instrument provides cleanest and largest blazar sample at present in the high energy range \citep{Ajello2022}. However, a large 
fraction of these gamma-ray sources remains unassociated with the blazar-subclasses and are referred to as the blazar candidates of uncertain type (BCUs) 
in the \emph{Fermi}-LAT point source catalogs. Observational identification of BCUs is a challenging task. In the last decade, machine learning and artificial 
intelligence based approaches have received enormous attention among the astrophysics community for astrophysical and cosmological studies 
\citep{Nun2014,Singh2019c,Baron2019,Singh2022a,Singh2022b} including the classification of gamma-ray sources with unknown 
nature \citep{Ackermann2012,Lee2012,Mirabal2012,Chiaro2016,Salvetti2017,Kova2019,Kova2020,Chiaro2021,Balakrishnan2021,Kaur2023}. 
\par
Here, we apply eXtreme Gradient Boosting (XGBoost) algorithm to the data set reported in \citep{Kaur2023} to identify the nature of a sample 
of 112 BCUs. This contains observational features of the sources in IR, optical, UV, X-ray and $\gamma$-ray wavebands. Therefore, 
it provides an opportunity to use the machine learning algorithms for better classification of BCU candidates using observed multi-wavelength 
information.   
In addition to this data set, we also use the redshift ($z$) measurements available in literature as an input to improve their identification. 
We start with a brief description of BCUs in the \emph{Fermi}-LAT era in Section \ref{BCUs-Fermi}. In Section \ref{data-set}, we describe the 
data set used in the present study. The XGBoost algorithm and its application in the identification of blazar subclasses are detailed in 
Section \ref{XGB}. We discuss the results in Section \ref{Res-Diss} and give the conclusions in Section \ref{Concl}.

\section{Blazar Candidates of Uncertain Nature in \emph{Fermi} Era}\label{BCUs-Fermi}
The \emph{Fermi}-LAT has been conducting a continuous monitoring of thousands of blazars for more than a decade. 
Based on the twelve years of science data collected by the \emph{Fermi}-LAT in energy range up to 1TeV, the most recent 
fourth source catalog (4FGL-DR3) reports more than 6600 point gamma-ray sources \citep{Abdollahi2022}. Of the 6658 sources in 
the 4FGL catalog, 389 are identified on the basis of pulsations, correlated variability with observations at other wavelengths, 
4112 have likely lower-energy counterparts, and the remaining 2157 sources remain unassociated. Blazars form the largest 
class in 4FGL catalog with 2251 associated with BL Lacs or FSRQs subclasses and 1493 as the BCUs. This suggests that 
BCUs represent a large fraction of the \emph{Fermi}-LAT blazars and call for new strategies to enable their proper 
identification and classification. BCUs include gamma-ray sources having low energy counterparts and showing blazar-like characteristics 
but their exact nature needs to be confirmed. Dedicated optical spectroscopic follow up campaigns are mainly required to identify the 
exact nature of BCUs observed by the \emph{Fermi}-LAT. However, large \emph{Fermi}-LAT positional uncertainty in the locations of 
gamma-ray sources as compared to the radio, optical and even soft X-ray observations prevent from the positional cross matching 
using standard astronomical procedures \citep{Massaro2013,Marchesini2019}. Therefore, data from the multi-wavelength observations 
along with the machine learning algorithms involving efficient statistical tools can be utilized to investigate the 
nature of BCUs \citep{Doert2013,Hassan2013,Salvetti2017,Kang2019,Chiaro2021}. 
		
\section{Data Selection}\label{data-set}
The main objective of present work is classification of a sample of 112 BCUs of the 4FGL catalog, reported by \cite{Kaur2023} (K23 hereafter), 
on the basis of broadband features using \emph{XGBoost} algorithm. K23 have used information derived from the multi-wavelength observations in the gamma-ray, X-ray, 
UV/optical and IR bands to classify 112 BCUs by applying neural networks. This data set is interesting because it contains information 
from both the gamma-ray observations with the \emph{Fermi}-LAT and the low energy counterparts. However, earlier classifications of BCUs of the 
4FGL catalog are based on the gamma-ray properties only \citep{Doert2013,Hassan2013,Salvetti2017,Kang2019,Chiaro2021}. Therefore, the unique broadband 
data set provided by K23 offers an opportunity to investigate the importance of multi-wavelength features in the classification of BCUs using efficient 
machine learning algorithms. Here, we briefly describe the data set reported by K23 and more details of the multi-wavelength data analysis and selection 
of 112 BCU candidates can be found in \citep{Kerby2021,Kaur2023}. Soft X-ray and UV/optical observations with the Neil Gehrels \emph{Swift} observatory have 
been used to identify likely counterparts in the sample of unassociated gamma-ray sources of the 4FGL catalog \citep{Kerby2021}. Out of 500 sources observed 
with the \emph{Swift}, authors found a total of 205 X-ray sources located within the 95$\%$ \emph{Fermi}-LAT positional uncertainty region of 188 unassociated 
sources in the 4FGL catalog. For such sources, a power law model was fit to derive the X-ray spectral indices and integral fluxes in the energy range of 0.3-10 keV. 
A photometric analysis of these source regions was also performed to obtain the magnitudes in the AB system. A neural network analysis predicted that out 
of 205 sources X-ray sources, 132 are blazar-like candidates with a probability (for a source being associated with a blazar class as defined by the underlying 
neural network approach) of more than 99$\%$. After employing the infrared properties in the $w1$, $w2$ and $w3$ bands from the WISE (Wide-field Infrared Survey Explorer) observations, a sample of 112 BCU candidates was obtained from the 4FGL catalog by K23.
\par
Following observed quantities of 1098 blazars of the 4FGL catalog, as defined and specified by K23, form the data set for this work. 
\begin{itemize}
	
    \item $\log{F_\gamma}$: $F_\gamma$ is the gamma-ray flux in units of $\rm{erg/s/cm^2}$ (\verb|Energy_Flux| in the 4FGL catalog)
    \item $\Gamma_\gamma$: Power law spectral index of gamma-ray photons (\verb|PL_Index| in the 4FGL catalog)
    \item \emph{Variability Index}: Year-over-year gamma-ray variability index (\verb|Variability_Index| in the 4FGL catalog)
    \item \emph{Curvature}: Significance of the curvature in the gamma-ray spectrum (\verb|PLEC_SigCurv| in the 4FGL catalog)
    \item $\log{F_X/F_\gamma}$: $F_X$ is the X-ray energy flux in the energy range 0.3-10keV 
    \item $\log{F_V/F_\gamma}$: $F_V$ is optical flux in the V-band 
    \item \verb|w1-w2|: Color index 1 measured by WISE
    \item \verb|w2-w3|: Color index 2 measured by WISE
\end{itemize}		

This basic data, designated as \emph{Set A} in the present study, is same as that used by K23. We also use the redshift ($z$) measurements 
as an additional information (not included by K23) to explore its importance in the classification of BCUs and define this data 
as \emph{Set B}. A brief description of these two data sets is given below:

\subsection{Set A}
\begin{figure}
\includegraphics[width=1.1\columnwidth]{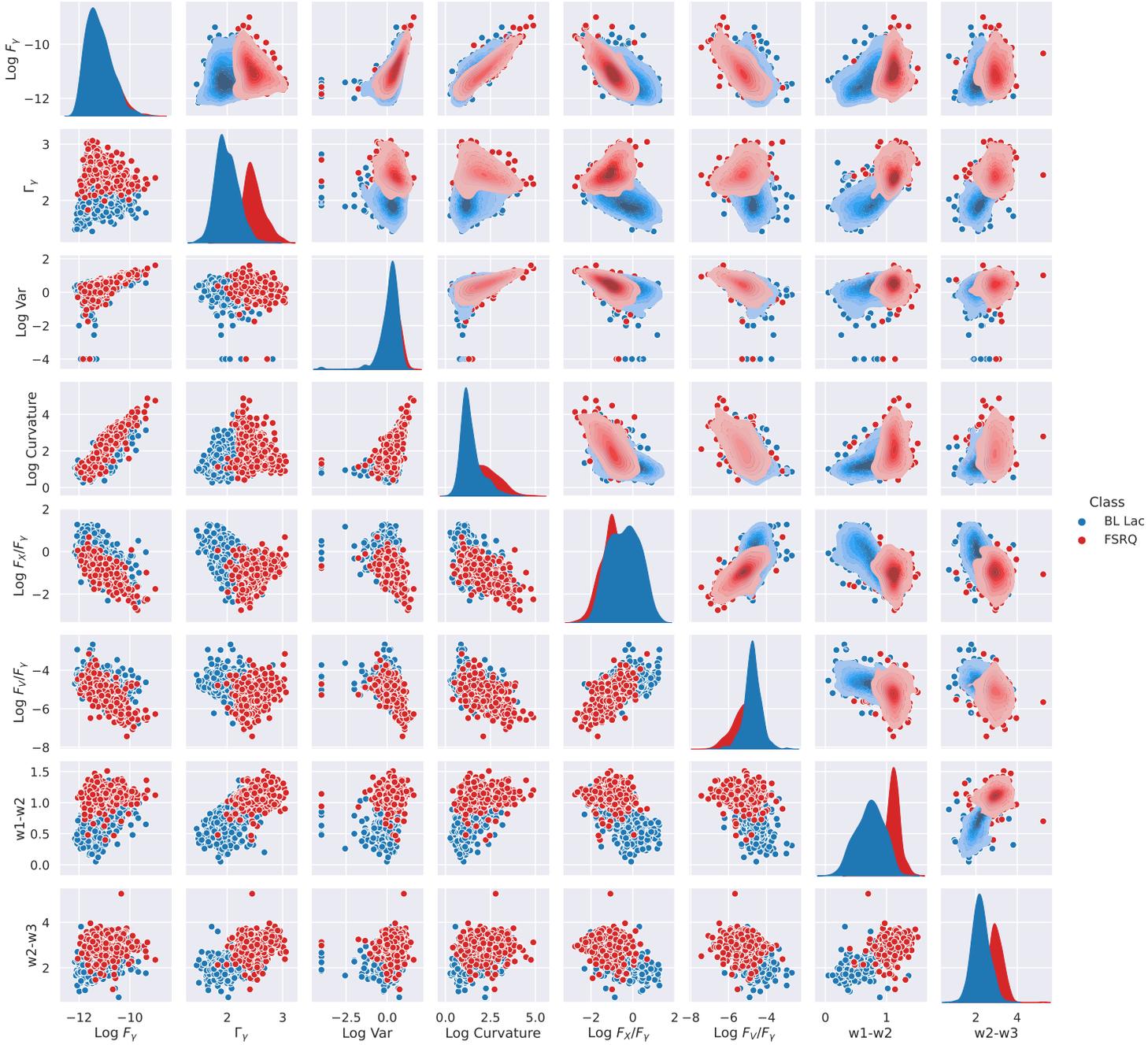}
\caption{Full pair plots of all the eight parameters described in Section \ref{data-set}. Blue and Red colors correspond to the  BL Lacs (671) and 
	FSRQs respectively. The diagonal plots are smoothed histogram distributions of parameters for each class of sources. Off-diagonal plots 
	in the upper and lower regions are the Kernel Density Estimate (KDE) and scatter plots for different parameters respectively.}
\label{fig:PPSetA}
\end{figure}

This data set consists of broadband features of 1098 known blazars (671 BL Lacs and 427 FSRQs) of the 4FGL catalog. The eight observed parameters 
(defined above) of 671 BL Lacs and 427 FSRQs of the 4FGL catalog are used for training and validation of the \emph{XGBoost} algorithm in this work. 
The frequency distributions and pair plots of these parameters for the BL Lacs and FSRQs used in training and validation are shown in Figure \ref{fig:PPSetA}. 
It is evident from Figure \ref{fig:PPSetA} that the parameter distributions, particularly the gamma-ray index ($\Gamma_\gamma$) and WISE color indices, 
of the BL Lacs and FSRQs are bimodal. Also, these two parameters, occupying different parameter spaces in correlation plots, seem to be the better features 
for identification of the blazar subclasses.

\subsection{Set B}
\begin{figure}
\includegraphics[width=1.1\columnwidth]{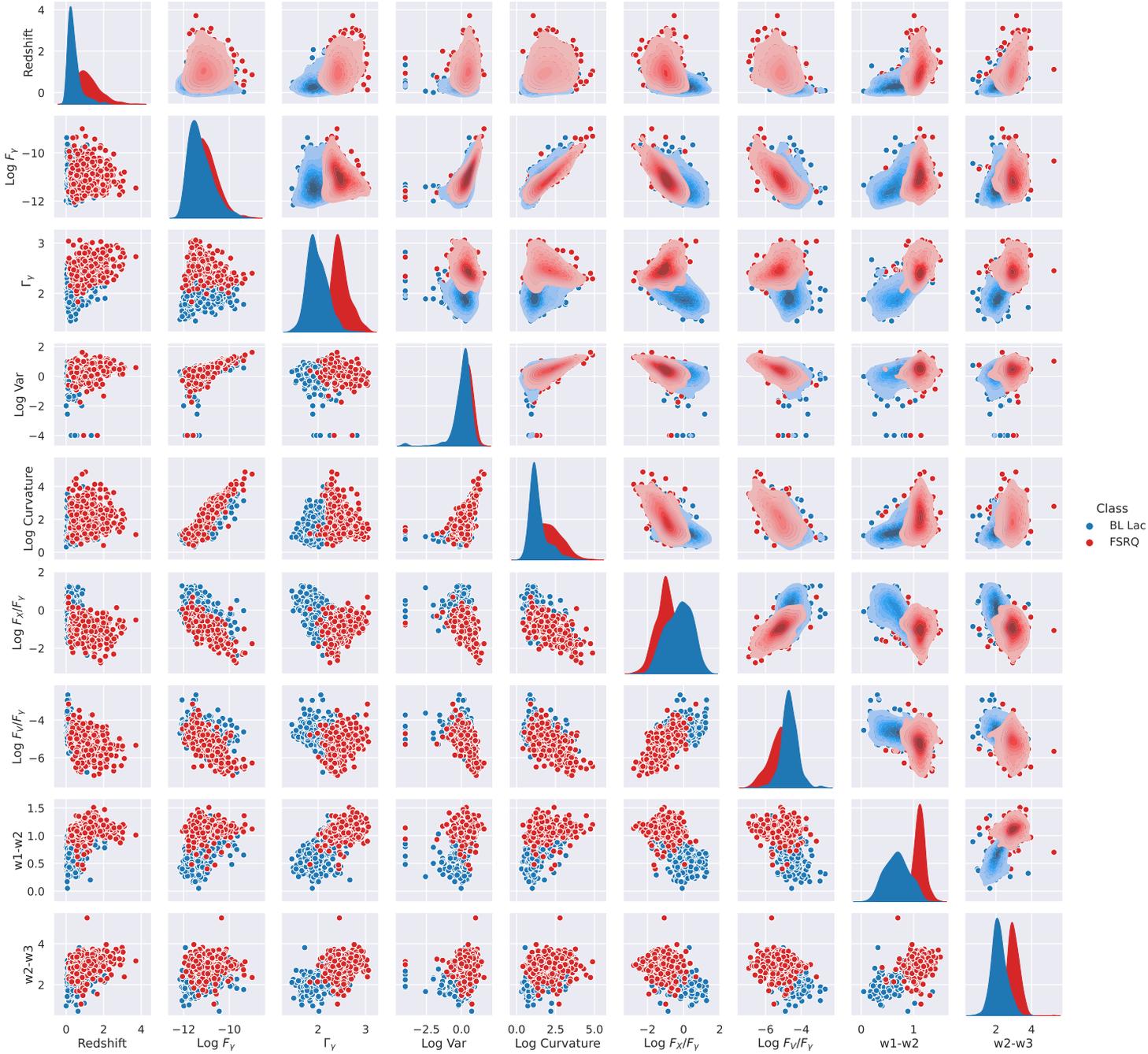}
\caption{Full pair plots of all the nine parameters described in \emph{Set B}. Blue and Red colors correspond to the  BL Lacs (479) and 
	FSRQs (407) respectively. The diagonal plots are smoothed histogram distributions of parameters for each class. The plots in the upper and 
	lower diagonal are the Kernel Density Estimate (KDE) and scatter plots for different parameters respectively.}
\label{fig:PPSetB}
\end{figure}

The redshift ($z$) distribution of FSRQs is significantly different from that of the BL Lacs \citep{Giommi2012,Ajello2022}. 
The mean redshift for FSRQs is $\sim$ 1.14, while for BL Lacs this is $\sim$ 0.55. Therefore, we want to explore whether redshift can also 
be used as a potential parameter in machine learning classification of gamma-ray sources. We have included the redshifts ($z$) of 
blazar sample in \emph{Set A} as an additional parameter and define this data sample as \emph{Set B}. It is derived from \emph{Set A} on the basis 
of the availability of redshift measurements of the sources. Out of 1098 blazars in \emph{Set A}, redshift measurements of 886 blazars are available 
in the 4th catalog of active galactic nuclei \citep{Ajello2020,Lott2020}. This sample of 886 blazars in \emph{Set B}, consisting of 479 BL Lacs and 
407 FSRQs, is used for training and validation purposes. Therefore, \emph{Set B} is a subset of \emph{Set A} and contains a total of nine 
input parameters for identification of the blazar subclasses. The distributions and pair plots of nine parameters for FSRQs and BL Lacs considered 
in \emph{Set B} are depicted in Figure \ref{fig:PPSetB}. 

\newpage
\onecolumn

\twocolumn
\section{Extreme Gradient Boosting (XGBoost)}\label{XGB}
In this work, we investigate the potential of \emph{eXtreme Gradient Boosting} (XGBoost) algorithm for classification of blazar 
candidates. It is a supervised machine learning algorithm which has been developed on the basis of \emph{Gradient Boosting 
Decision Tree} (GBDT), also known as gradient boosting machine or gradient boosted regression tree \citep{Friedman2001,Chen2016}.
XGBoost classifier, a scalable machine learning system, has appeared more recently and is able to provide 
state-of-the-art results with stability on a wide range of problems in the fields of machine learning and data 
mining challenges. In recent times, it has been used in astronomy for different purposes including classification of 
gamma-ray sources detected by the \emph{Fermi}-LAT \citep{Li2019,Li2021,Ivanov2021,Agarwal2023,Sahakyan2023}.  
The scalability of XGBoost is attributed to several innovative algorithmic optimizations 
including a novel tree learning algorithm to handle the sparse data and a theoretically justified weighted
quantile sketch procedure for handling instance weights in approximate tree learning \citep{Chen2016}. 
\par
Similar to the GBDT, the XGBoost algorithm applies boosting method and builds strong classifiers by means of 
learning multiple weak classifiers sequentially. However the XGBoost excels in performance and efficiency and also 
differs from the GBDT in the calculation of residuals. Both the first and second derivatives of the loss function 
are used for calculating the residuals in the XGBoost whereas only first derivative is used in the case of GBDT. 
The XGBoost model typically consists of a certain number of trees. The first tree is constructed after applying 
a cost function which tests all features and chooses the best features and conditions from data. 
Specifically, the feature with the highest gain is chosen as the root node, and its corresponding condition
is decided during the first epoch of testing. Subsequently, features are chosen and added to the tree as 
internal nodes one by one recursively until the tree grows to the specified depth or the maximum depth. 
The ends of branches are called leaves and they hold the results of the tree. The results in leaf nodes decide 
degree of an unknown source to belong to a particular class. Similarly, the $(i+1)^{th}$  tree is built in the 
same way as the first tree, except that the aim of this tree is to fit the residuals i.e. the measurement of the
difference between the predicted and the original target of the previous model consisting of $i$-trees rather 
than the original target. The residuals are fitted with a certain learning rate which is related to the step 
size of the fitting. Finally, the process of building the XGBoost classifier stops when a specific number of 
trees or number of estimators has been built. In the present work, we have used XGBoost python package provided by 
\emph{scikit-learn} as an open source \citep{Pedregosa2011}. Application of \emph{XGBoost} machine learning algorithm for 
identifying the exact nature of BCUs is described below.

\subsection{Data Balancing with SMOTE}
\begin{figure}
 	\includegraphics[width=\columnwidth]{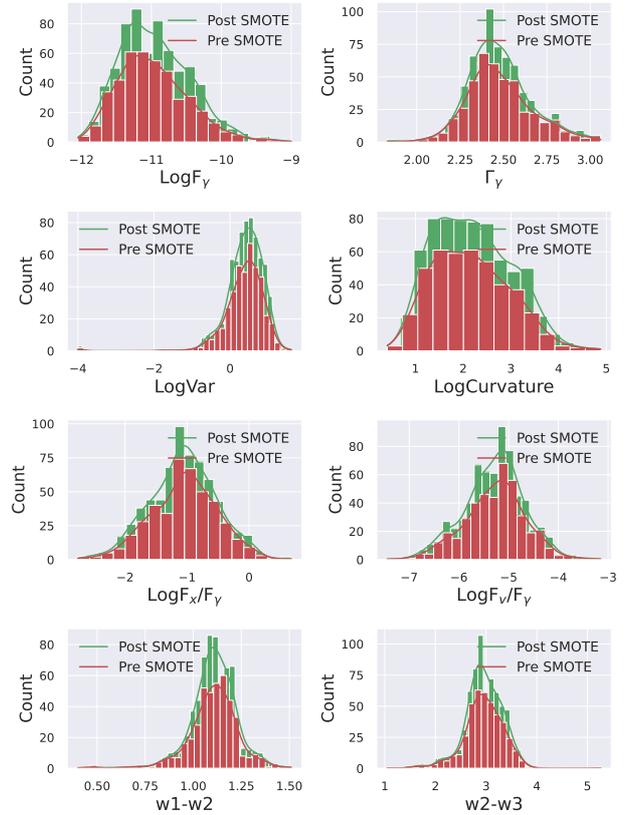}
	\caption{Smoothed distributions of the eight parameters of the FSRQs before and after employing SMOTE for balancing 
		 the data in \emph{Set A}. The two distributions for all the parameters are almost identical and hence the 
		 synthetically generated FSRQs follow the distributions of original FSRQs.}
 \label{fig:SMOTE-A}
\end{figure}

\begin{figure}
	\includegraphics[width=\columnwidth]{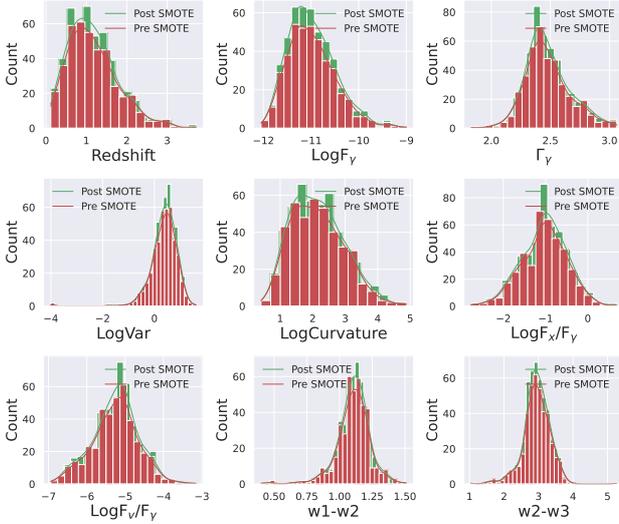}
 	\caption{Smoothed distributions of the nine parameters of the FSRQs before and after the application of SMOTE for data balancing in 
	         in \emph{Set B}. Here also, the two distributions are identical.}
	\label{fig:SMOTE-B}
\end{figure}

The Synthetic Minority Over-sampling Technique (SMOTE) is used in astronomy to deal with the imbalanced 
data sets \citep{Bethapudi2018,Sutrisno2019,Carruba2023}. It has also been used for classification of gamma-ray 
sources \citep{Kaur2019,Kaur2021,Kaur2023}. In the present work, the number of BL Lacs is more than that of the FSRQs 
in both data \emph{Set A \& B} (Section \ref{data-set}). 
Thus, the two blazar subclasses are not equally represented in the data sets for training the XGBoost algorithm. 
Such imbalance in the data leads to the selection effects in the ensemble based algorithms which may be biased 
against the under represented classes \citep{Last2017}. In order to balance the  representation of BL Lacs and 
FSRQs in the data \emph{Set A \& B}, we employ Synthetic Minority Over-sampling Technique (SMOTE) for generating  
additional synthetic members of the under represented blazar subclass (i.e. FSRQ) using  k-nearest neighbor 
approach \citep{Chawla2002}. As a result of the SMOTE expansion, the data \emph{Set A \& B} now contain original 
BL Lacs and FSRQs as well as new synthetic FSRQs generated with the same distribution in parameter-space as the real FSRQs. 
Thus, the training data samples in \emph{Set A} and \emph{Set B} are balanced with a total of 671 FSRQs/BL Lacs and 
479 FSRQs/BL Lacs respectively. A comparison of the Pre-SMOTE and Post-SMOTE smoothed distributions of each parameter 
for FSRQs in \emph{Set A} and \emph{Set B} is presented in Figures \ref{fig:SMOTE-A} and \ref{fig:SMOTE-B} respectively. 
It is observed that changes in distributions of the parameters are minimal and the artificially generated FSRQs mimic 
the original FSRQs in both the data sets. These balanced data sets are used for training the XGBoost algorithm to 
classify the BCUs among FSRQs and BL Lacs.

\subsection{Training and Validation}\label{Train-Valid}
We use the blazar data samples described in \emph{Set A} and \emph{Set B}, with eight and nine input parameters respectively, for 
training and validation of the XGBoost algorithm in the present work. We define two \emph{XGBoost} models namely XGB and XGBz corresponding to 
the \emph{Set A} and \emph{Set B} respectively. The main difference between the two models is the number of input parameters to each model.  
We randomly split each data set into training and validation data sets in 80:20 ratio so that the two data samples are balanced.
Thus, out of 1342 blazars (original$+$SMOTE) in \emph{Set A}, 1073 blazars are randomly selected for training and remaining 269 blazars 
for validation of the XGB model. Similarly in case of \emph{Set B}, 766 and 192 blazars amongst the total 958 blazar (original$+$SMOTE) 
samples are reserved for training and validation of the XGBz model. Application of trained model on the unassociated sources in the 
validation data set returns a set of probability values referred to as \verb|FSRQness| for each source. The \verb|FSRQness| value indicates 
probability of a source likely to be an FSRQ. For the predictions, the evaluation regards the instances with \verb|FSRQness| value greater 
than 0.5 as FSRQs and others as BL Lacs. We optimize the model parameters by minimizing the binary classification error rates on the 
validation data sets for both the models. The classification \emph{error rate} is calculated as 
\begin{equation}\label{eqn:error}
	\rm Error~rate = \frac{\rm number~of~wrong~predictions}{\rm total~number~of~predictions}
\end{equation}
We monitor the performance of a model on each training epoch by evaluating the classification error rate on both training as well as validation 
data sets. As the training progresses on different epochs, classification error rate continuously decreases for the training data set as 
the model learning improves. However, for validation data the error rate first decreases as the model gets trained and after certain 
epochs the error rate remain same and then it starts increasing as the model overfits the training data. We stop the training when the 
error rate does not change since further training will have negative impact on the predictive capabilities of the model on unknown data.
The final XGB model consists of 380 trees (total number of estimators), each having depth of 3 (maximum depth) and a learning rate of 0.1.
Whereas, the optimized XGBz model has 535 trees, each with depth 2 and a learning rate of 0.1. All other hyper-parameters of the model are 
set to their default values.
\subsection{Model Performance Evaluation}
\begin{figure}
\includegraphics[width=1\linewidth]{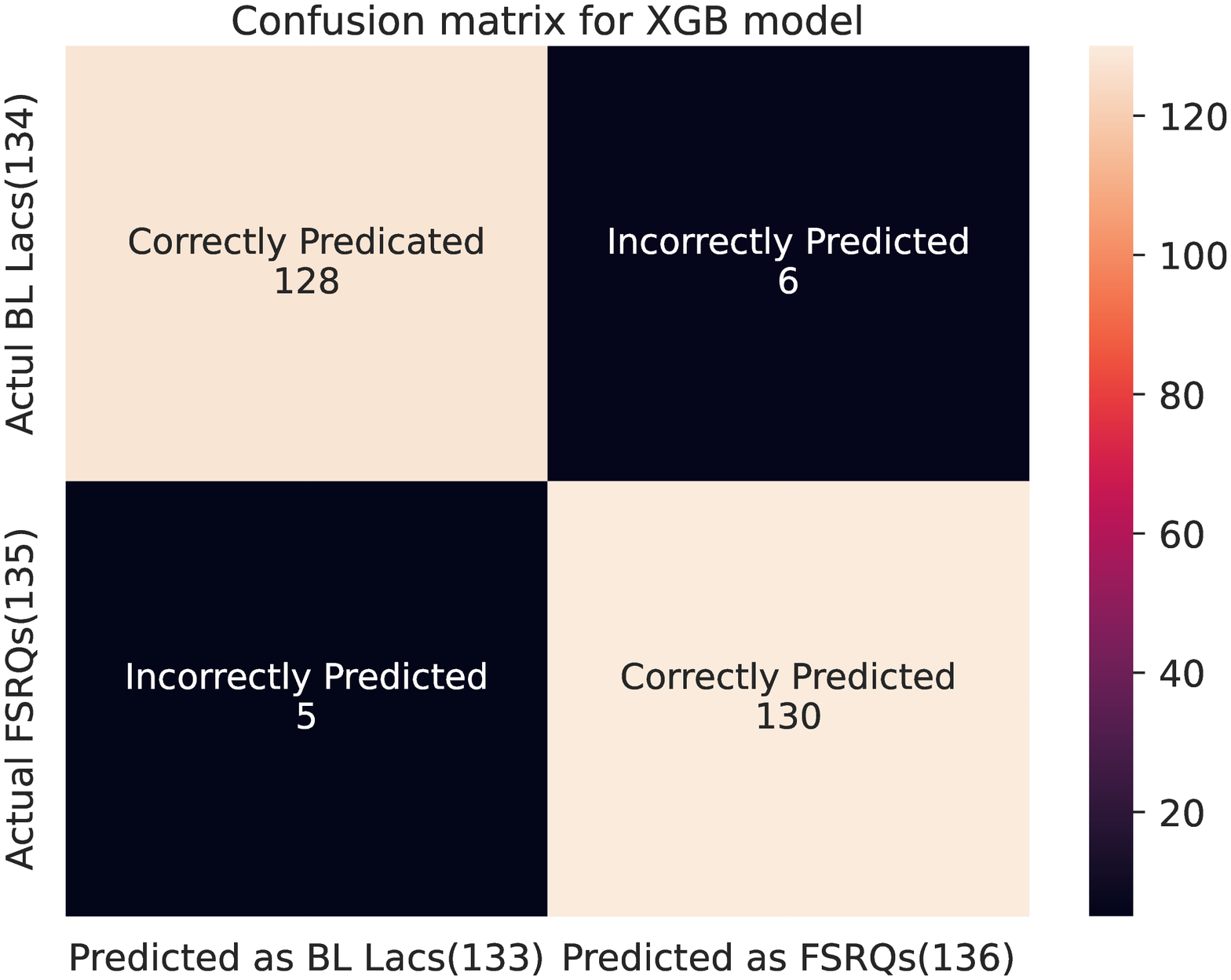}
\includegraphics[width=1\linewidth]{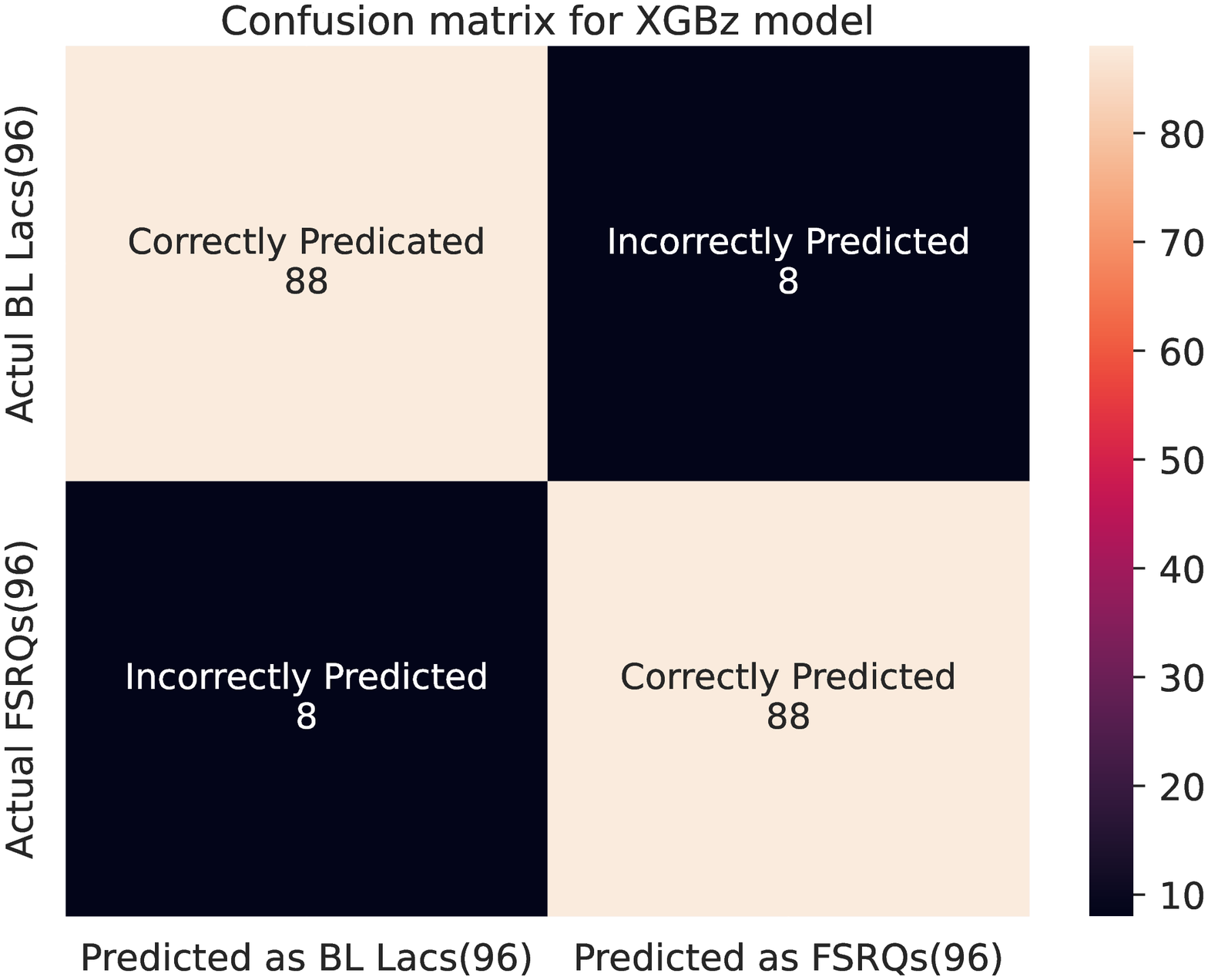}
\caption{Confusion matrices for the XGBoost classifiers XGB and XGBz on 269 and 192 known sources respectively.}
\label{fig:conf-matrices}
\end{figure}

In machine learning based studies, evaluation metrics are generally used to quantify the the performance of an algorithm or classifier.
We apply the trained models XGB and XGBz on validation data of \emph{Set A} and \emph{Set B} respectively to calculate three standard 
metrics corresponding to the accuracy, precision and recall to determine model performance. \emph{Accuracy} is defined as the 
fraction of total number of correct predictions among the total number of predictions. \emph{Precision} is the ratio of true positive 
predictions to all the positively predicted examples. \emph{Recall} is the ratio of true positive predictions to all the true 
positive examples. Qualitatively, \emph{Recall} is a measure of sensitivity of a model to a particular class and it represents 
the capability of a model to find all the relevant cases within a data set whereas \emph{Precision} is the ability of a classifier 
to identify only the relevant data points. Precision and Recall are defined for each class, whereas the Accuracy is defined for 
whole data set. A summary of the performance evaluation of the two models using \emph{Accuracy}, \emph{Precision} and \emph{Recall} 
metrics is provided in Table \ref{tab:model-performance}. It is observed that the XGB and XGBz models predict class of the unassociated 
sources with an accuracy of 96$\%$ and 92$\%$ respectively. The confusion matrices for both the models are also shown in 
Figure \ref{fig:conf-matrices}. The confusion matrix is a specific table layout that summarizes the prediction results of a classification 
model in machine learning. It describes the data in a matrix form according to two criteria namely the real category and the prediction 
judgement made by the classification model.

\section{Results and Discussion}\label{Res-Diss}
Machine learning algorithms are used as probabilistic classifiers of unassociated source. In this work, we 
apply XGBoost method, a supervised machine learning algorithm, to identify the BL Lacs and FSRQ  subclasses 
among the sample of selected blazars of uncertian type. Main results of this study are discussed below:

\subsection{Classification of BCUs}
We employ the trained models XGB and XGBz (subsection \ref{Train-Valid}) on the sample of BCUs in \emph{Set A} 
and \emph{Set B} respectively. During validation of the algorithm, we have classified sources to be \emph{likely FSRQ} 
if \verb|FSRQness| $>$ 0.5 and \emph{likely BL Lacs} if \verb|FSRQness| $<$ 0.5. But we adopt a more strict and conservative 
approach while classifying the BCUs. We define sources to be \verb|highly likely FSRQ| if \verb|FSRQness| $>$ 0.99, 
\verb|highly likely BL Lacs| if \verb|FSRQness| $<$ 0.01 and remaining as \verb|ambiguous-type blazars|. 
The associated class and corresponding \verb|FSRQness| value for each source classified in the present work using the XGB and XGBz models 
along with the corresponding results reported by K23 are summarized in Tables \ref{tab:classification-Res1}, and \ref{tab:classification-Res2}. 
Results obtained in the present work with data \emph{Set A and B} are discussed below:

\subsubsection{Set A}
\begin{figure}
\includegraphics[width=\columnwidth]{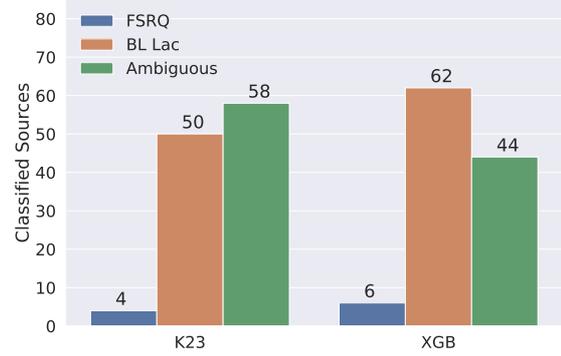}
\caption{Comparison of the classification of 112 BCUs using XGB model with the results reported by K23 \citep{Kaur2023}.}
\label{fig:Classification-Hist-XGB} 
\end{figure}

\begin{figure}
\includegraphics[width=\columnwidth]{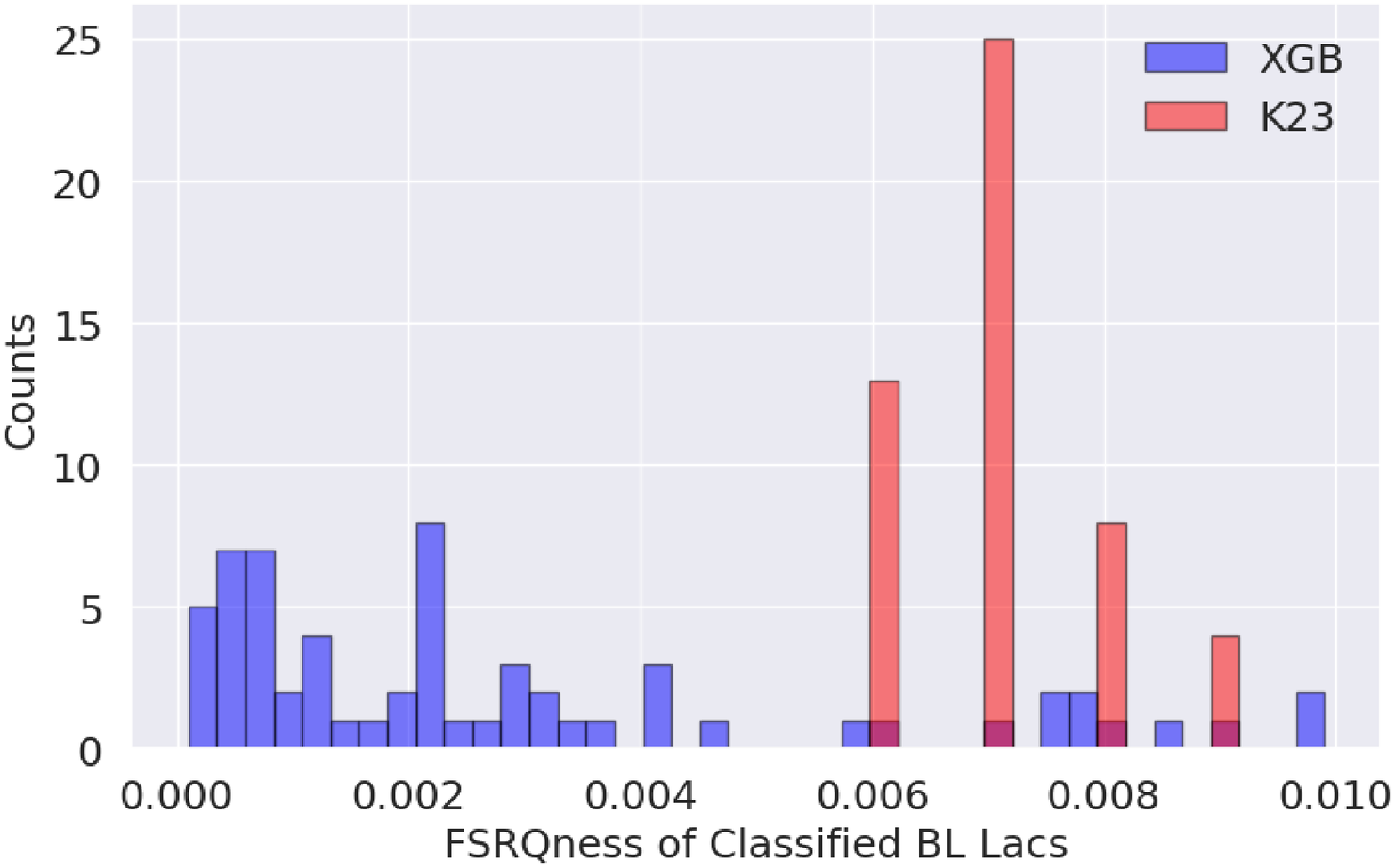}
\includegraphics[width=\columnwidth]{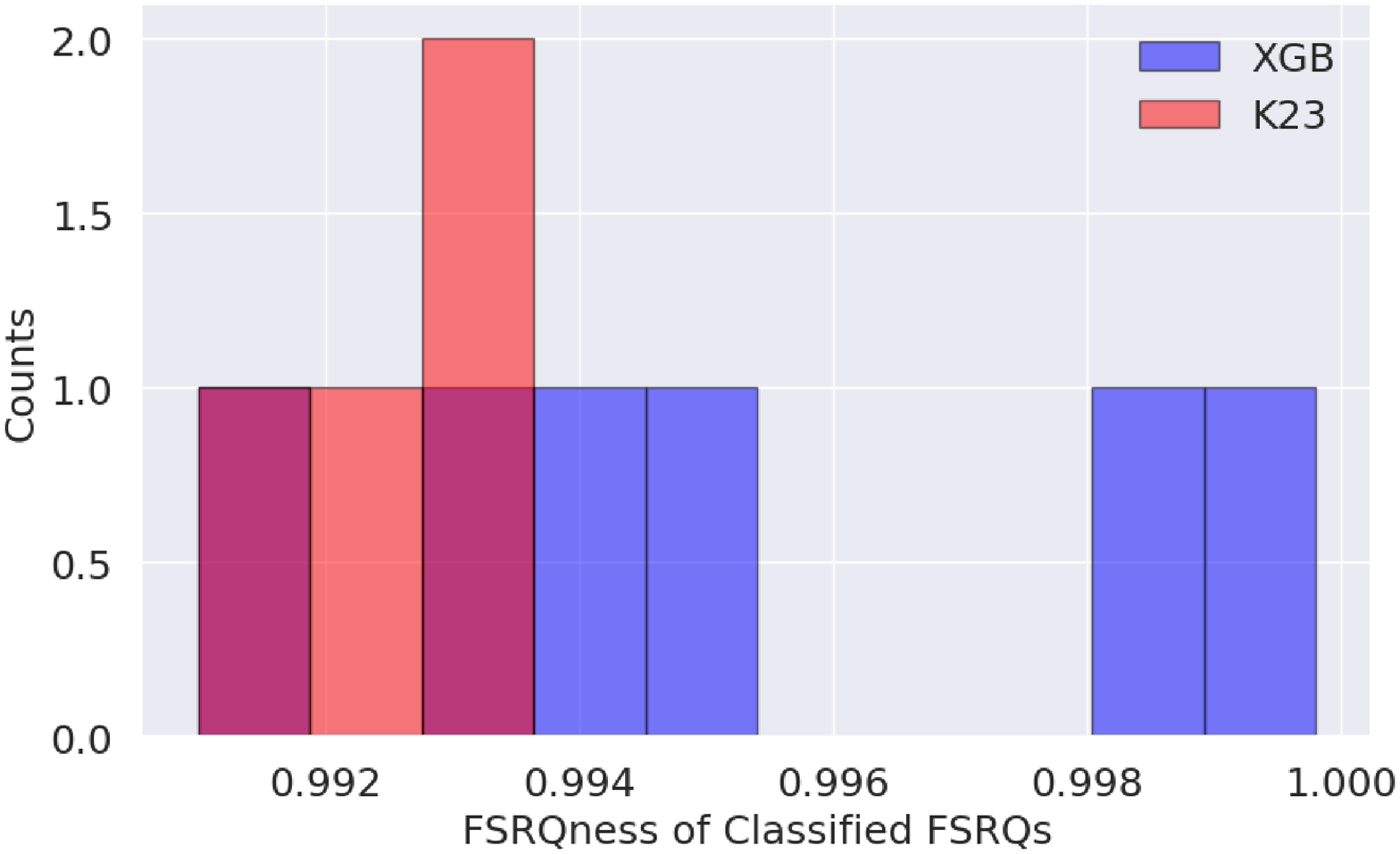}
\caption{Distribution of FSRQness values of classified BL Lacs and FSRQs using XGB model and results reported by K23 \citep{Kaur2023}.}
\label{fig:FSRQness-Compare-XGB-K23}
\end{figure}

There are 112 BCUs in \emph{Set A} for classification. Application of XGB model on this data sample results in the identification of 6 
sources as \verb|highly likely FSRQ|, 62 sources as \verb|highly likely BL Lacs| and remaining 44 sources as \verb|ambiguous-type blazars|. 
These numbers indicate a significant improvement in the earlier classification of the same sample using the neural networks reported by K23 
\citep{Kaur2023} wherein 4 sources are identified as \verb|highly likely FSRQs|, 50 sources as \verb|highly likely BL Lacs| and remaining 
58 sources as \verb|ambiguous-type blazars|. A comparison of the performance of XGB model with that of the neural networks (K23) in terms of 
the number of sources is shown in Figure \ref{fig:Classification-Hist-XGB}. 
The distributions of FSRQness values of sources classified as BL Lacs and FSRQs using XGB model and neural networks (K23) are shown 
in Figure \ref{fig:FSRQness-Compare-XGB-K23}. It is observed that the mean of FSRQness values obtained from XGB model for classified BL Lacs 
(FSRQs) is closer to 0 (1) than FSRQness values of classified BL Lacs (FSRQs) reported by K23. With XGB model, the mean of FSRQness values of 
classified  BL Lacs (FSRQs) is 0.0029 (0.9952), whereas with the neural network methods the mean of FSRQness values of classified BL Lacs (FSRQs) is 
0.0071 (0.9923). This indicates that \emph{XGBoost} based XGB model classifies the sources with more confidence/decisiveness by associating relatively 
higher probabilities (for belonging to a particular class) to the sources compared to neural networks based algorithm reported by K23. Therefore, 
the results obtained in the present work suggest that the \emph{XGBoost} algorithm classifies the sample of 112 BCUs better than the neural network based 
methods.

\subsubsection{Set B}
\begin{figure}
\includegraphics[width=\columnwidth]{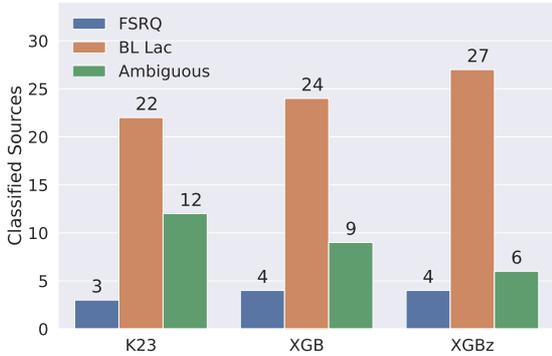}
\caption{Comparison of the classification of 37 BCUs with known redshift using XGBz model with the XGB model and results reported by K23 \citep{Kaur2023}.}
\label{fig:Classification-Hist-XGBz}
\end{figure}

\begin{figure}
\includegraphics[width=\columnwidth]{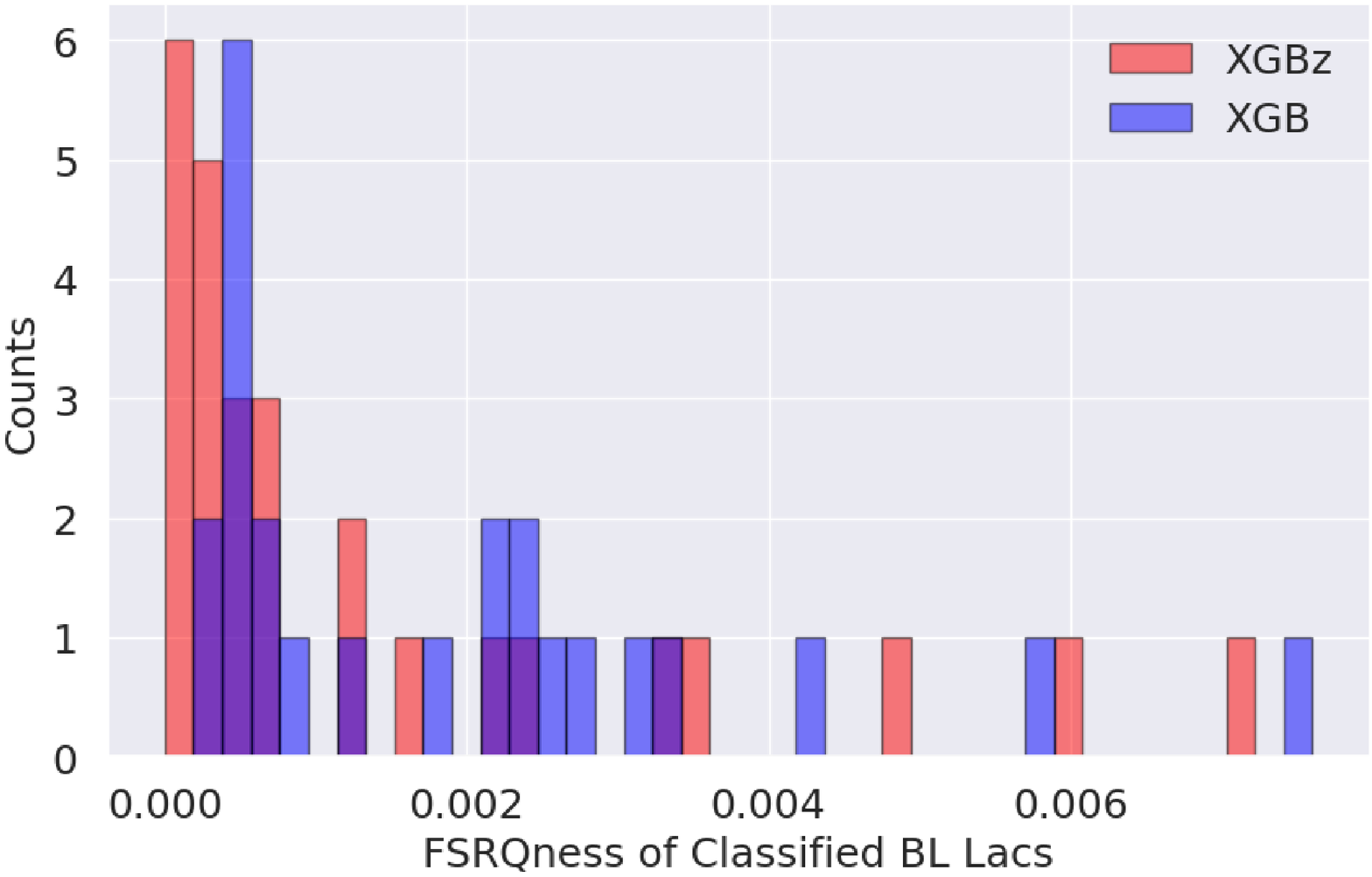}
\includegraphics[width=\columnwidth]{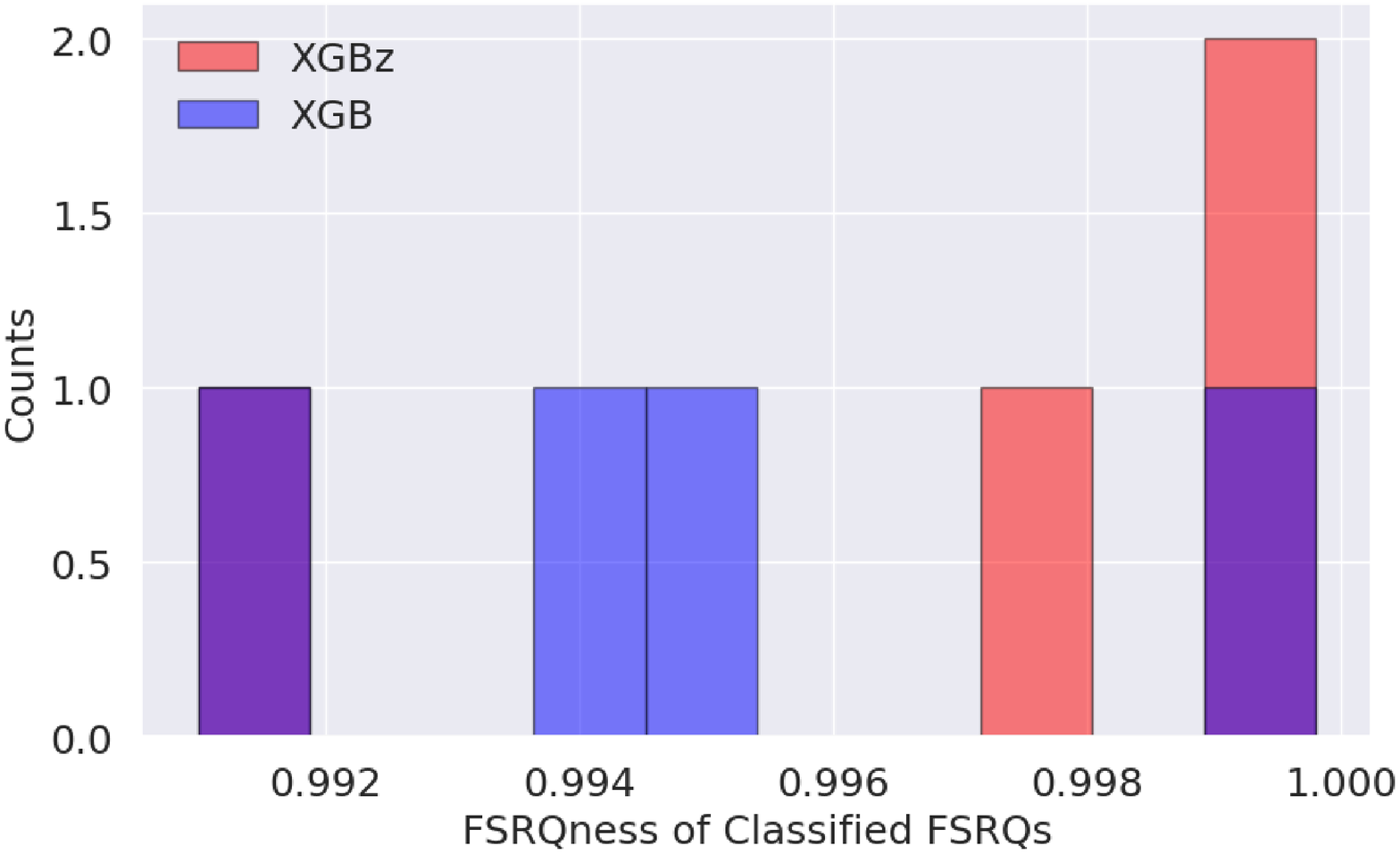}
\caption{Distribution of FSRQness values of classified BL Lacs and FSRQs using XGB and XGBz model.}
\label{fig:FSRQness-Compare-XGBz-XGB}
\end{figure}

Out of 112 BCUs in \emph{Set A}, redshifts of only 37 sources are available for classification in \emph{Set B}. Application of XGBz model on the 
sample of 37 BCUs classifies 4 sources as \verb|highly likely FSRQs|, 27 sources as \verb|highly likely BL Lacs| and 6 sources as \verb|ambiguous-type blazars|. 
This is compared with the output of XGB model and results reported by K23 without using $z$ information in Figure \ref{fig:Classification-Hist-XGBz}. The 
XGB model (K23) classification of these 37 BCUs results into 4(3) as \verb|highly likely FSRQs|, 24(22) \verb|highly likely BL Lacs| and 9(12) 
as \verb|ambigous-type blazars|. 
This implies that the XGBz model provides a better classification than the XGB model and the neural network approach. The 
list of sources that change the classification in XGBz model compared to XGB model are tabulated in Table \ref{tab:z_Compare}. The 6 ambiguous XGB sources 
are reclassified as FSRQs/BL Lacs by the XGBz model with 8.73$\%$ change in mean FSRQness. However, 3 classified XGB sources are identified as ambiguous by 
XGBz model with 1.4$\%$ change in mean FSRQness. Effectively, 3 additional sources are classified by the XGBz as compared to XGB model. But the data sample 
of 37 BCUs is not significant enough to establish the robust impact of redhsift in the classification of these sources. Figure \ref{fig:FSRQness-Compare-XGBz-XGB} 
shows the distributions of FSRQness values of sources classified as BL Lacs and FSRQs using XGB and XGBz models. This suggests that the mean of FSRQness 
values obtained from XGBz model for classified BL Lacs (FSRQs) is slightly close to 0 (1) as compared to FSRQness values of classified BL Lacs (FSRQs) obtained 
from XGB model. Using XGBz model, the obtained mean of FSRQness values of classified  BL Lacs (FSRQs) is 0.0014 (0.9968), whereas with XGB model the mean of 
FSRQness values of classified BL Lacs (FSRQs) is 0.002 (0.9950). Therefore, use of redshift as an additional parameter in XGBz model has an impact on the 
classification, but more number of BCUs with redshift information is required to comment on its significance and robustness.

\subsection{Feature Importance}
\begin{figure}
\includegraphics[width=\columnwidth]{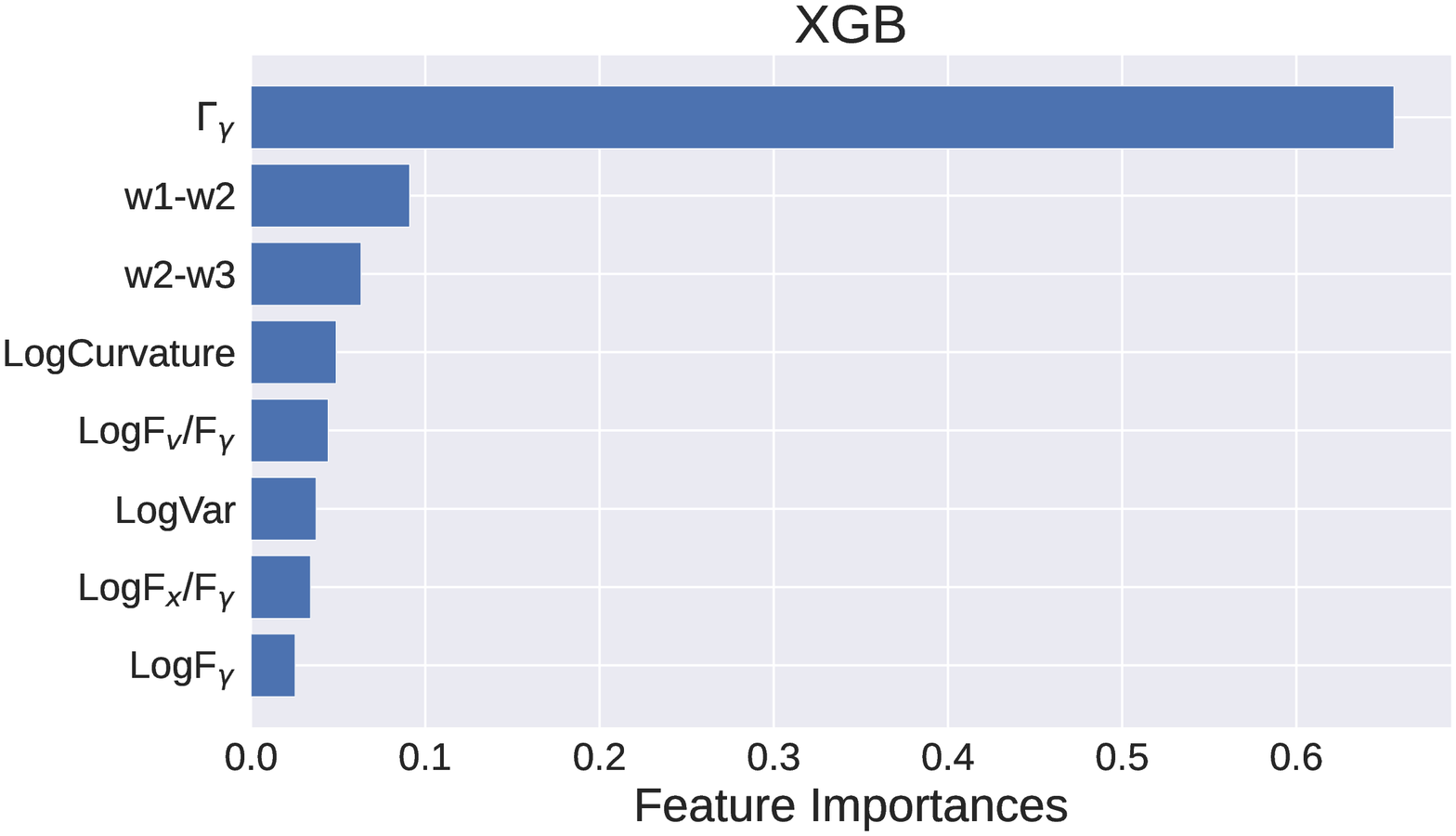}
\includegraphics[width=\columnwidth]{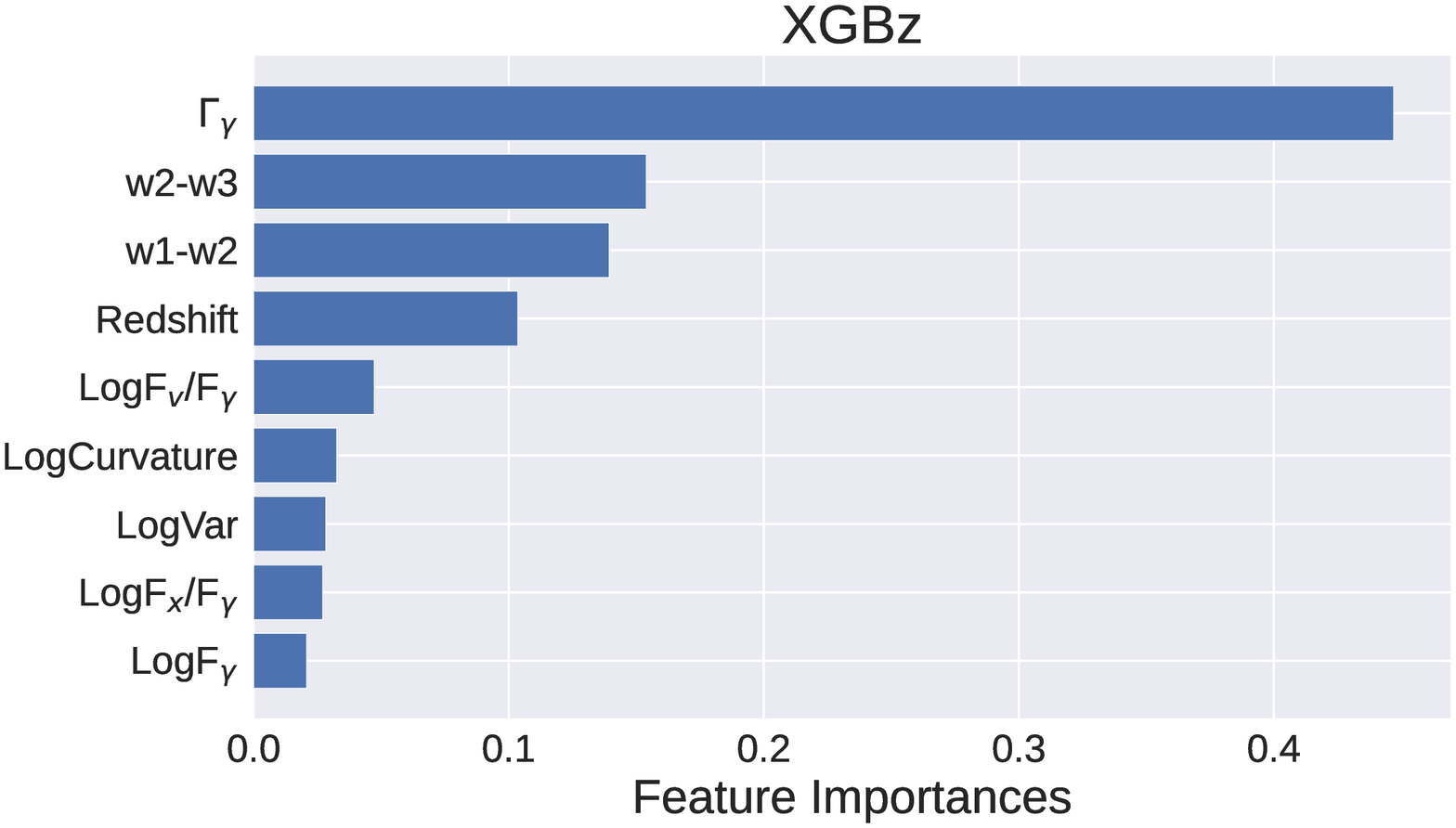}
\caption{Relative importance of different input parameters or variables in the XGB and XGBz models for classification of blazar candidates of uncertain type.}
\label{fig:Feature-Imp}
\end{figure}

In machine learning algorithms, feature importance is estimated to rank the importance of each input feature/parameter. It basically 
represents the relative weight of the variables and indicates how useful or valuable each feature is in the construction of the boosted 
decision trees within the model. The more a feature is used to make key decisions with decision trees, its relative importance will be higher.
This is explicitly calculated for each feature in the in the data set, allowing attributes to be ranked in comparison to each other. 
Importance is calculated for a single decision tree by the amount each attribute split point improves the performance measure, weighted by the number of 
observations the node is responsible for. The feature importances are then averaged across all the decision trees within the model \citep{Hastie2016}. 
We have calculated the feature importances of each attribute in the data \emph{Sets A and B} described in Section \ref{data-set} for the models XGB 
and XGBz respectively. The same are plotted in Figure \ref{fig:Feature-Imp} for the two models used in this work. 
It is observed that the gamma-ray photon index parameter ($\Gamma_\gamma$; \verb|PL_Index|) is the dominant feature in deciding the subclass of blazars 
for both the models. This is in agreement with the distribution plots in Figure \ref{fig:PPSetA} wherein $\Gamma_\gamma$-distributions for BL Lacs and 
FSRQs are nicely separated. The WISE infra-red color indices (\verb|w1-w2| and  \verb|w2-w3|) are the second and third most dominant features in the 
classification. Therefore IR data play an important role in classifying the BCUs. Classification of blazars, based on the observational features, offers 
unique opportunity to identify key characteristics and probe the physical processes involved in their broadband emissions. For example, IR measurements 
involve contributions from the jet and molecular torus for FSRQs whereas gamma-ray observations only probe the non-thermal emission of the beamed jet. 
FSRQs usually have a steep gamma-ray spectrum in the GeV-TeV domain while BL Lacs (mainly HBLs) show hard spectrum \citep{Prandini2022}. 
\par
As discussed earlier, the main difference between XGB and XGBz models is the use of redshift ($z$) information as an additional feature. 
From Figure \ref{fig:Feature-Imp}, it is evident that $z$ is the fourth most dominant parameter in deciding the class of BCUs using the XGBz model. The 
XGBz model has performed better than the XGB model during the classification of a sample of 37 BCUs mainly because of the use of redshift information. 
Even though redshift has played a role in classification, but it is not the dominant feature. We need a large sample of BCUs with known redshift to 
establish the explicit role of redshift in the identification of blazar subclasses more robustly. Also distinct and well separated distributions of redshift 
for BL Lacs and FSRQs in Figure \ref{fig:PPSetB} reflect a bias in observation of two types of blazars. Since BL Lacs have weak or no spectral lines, their 
redshift measurement is a challenge as compared to FSRQs.
\section{Summary}\label{Concl}
The present work underscores the potential of XGBoost algorithm for the classification of blazar candidates 
of uncertain type and compares its performance with the neural network classifiers. Results obtained in this 
work indicate a significant improvement in the classification of 112 candidates as compared to the neural networks based 
approach of K23 \cite{Kaur2023}. The main findings of this study are summarized below:
\begin{itemize}
	\item The gradient boosting classifier based on the XGBoost implementations is found to be a data efficient 
		alternative to the neural network based classifiers. Also, it is relatively straightforward to retrieve the 
		importance scores for each attribute after the decision trees are constructed in the gradient boosting implementations. 
		The XGBoost algorithm is able to better classify the sources with more robustness/decisiveness than neural networks.
		Efficacy and accuracy of the XGBoost classifier may outperform the other machine learning based classifiers 
	       in the identification of the exact nature of blazar candidates of uncertain type.

       \item  Out of 112 blazar candidates of uncertain type, the \emph{XGBoost} classifier is able to identify 6 sources as FSRQ and 
	      62 objects as BL Lacs with very high accuracy using only multi-wavelength information. The mean of FSRQness values 
		of classified  BL Lacs (FSRQs) is 0.0029 (0.9952), whereas with the neural network methods the mean of FSRQness values of 
		classified BL Lacs (FSRQs) is 0.0071 (0.9923).

       \item  Among the eight observational features used for classification, gamma-ray spectral index and IR color indices play 
	      dominant role in identifying the blazar subclasses among the  blazar candidates of uncertain type.

       \item Use of redshift as an additional feature leads to the identification of 4 sources as FSRQs and 27 sources as BL Lacs 
	     in the sample of 37 BCUs. Without using redshift information, only 24 sources are classified as BL Lacs in the same sample 
	     of 37 BCUs whereas the identification of FSRQs is unchanged.	
		
     \item  Redshift of the blazars is found to play a role  in the classification of the blazar candidates of uncertain type. It is 
	     supported by the observational fact that BL Lac type blazars are located at smaller redshift and FSRQs are mostly located at high 
	     redshifts. But a more rigorous investigation of a large sample of BCUs with known redshift is needed to establish the importance 
		of redshift in blazar classification since there is a bias in the redshift measurements of FSRQs as compared to BL Lacs.  
\end{itemize}		
Over the last two decades, rapid development in field of ground-based gamma-ray astronomy, using state-of-the-art imaging atmospheric Cherenkov 
telescopes, has provided detection of more than 80 blazars in the GeV-TeV energy range. Among these blazars, about 70 have a measured/known redshifts. 
This emerging population of TeV blazars offers a new perspective to be investigated in future as a large fraction of BCUs detected by the \emph{Fermi}-LAT 
still remains unassociated. The information derived from the ground-based observations such as photon spectral index in the TeV energy range along 
with the redshift will be more effective in the classification of the blazar candidates of uncertain type using the potential of XGBoost machine 
learning algorithm explored in the present work.  

\section*{Data Availability}
Data used in this work is derived from the literature and can be shared on reasonable request.
\section*{Acknowledgements}
Authors thank the anonymous reviewer for her/his critical and valuable suggestions to improve the contents of the manuscript.
\newpage
\onecolumn
\begin{table}
\begin{center}
\caption{Summary of the performance evaluation of the XGBoost classifiers XGB and XGBz.}
\begin{tabular}{rccccccc}
\hline
	&&&&&&&\\
	Model    &\multicolumn{2}{c}{Data sample} &Model-Accuracy(\%) &\multicolumn{2}{c}{BL Lac} &\multicolumn{2}{c}{FSRQ}\\
	&&&&&&&\\
\hline
	&&&&&&&\\
	 	&Training  	&Validation  & &Precision(\%) &Recall(\%) &Precision(\%) &Recall(\%)\\
	&&&&&&&\\
\hline
	&&&&&&&\\
	XGB   &1073  &269 &96  &96  &96   &96  &96   \\
	XGBz  &766   &192 &92  &92  &92   &92  &92   \\
	&&&&&&&\\
\hline
\end{tabular}
	\label{tab:model-performance}
\end{center}
\end{table}
\begin{table}
\begin{center}
\caption{Summary of results for the classification of 112 BCUs using XGBoost algorithm.}
\begin{tabular}{rccccccc}
\hline
	&&&&&&\\
	    \multicolumn{2}{c}{Source Name}  & \multicolumn{2}{c}{K23}        &\multicolumn{2}{c}{Present work} &\multicolumn{2}{c}{Present work}\\
	    \multicolumn{2}{c}{}  & \multicolumn{2}{c}{}        &\multicolumn{2}{c}{XGB Model}&\multicolumn{2}{c}{XGBz model}\\
	&&&&&&&\\
\hline
	&&&&&&&\\
	4FGL Name & UVOT Name & Assigned class  &FSRQness  & Assigned class& FSRQness& Assigned Class& FSRQness\\
	&&&&&&\\
\hline

	&&&&&&\\
J1008.2-1000&J100802.5-095918&FSRQ&0.993&FSRQ&0.9937&FSRQ&0.9995\\
J1008.2-1000&J100749.4-094912&FSRQ&0.993&Ambiguous&0.9026&-&-\\
J1637.5+3005&J163728.1+300953&FSRQ&0.992&Ambiguous&0.9468&FSRQ&0.9972\\
J1637.5+3005&J163739.2+301013&FSRQ&0.991&FSRQ&0.9951&Ambiguous&0.9747\\
J0026.1-0732&J002611.6-073115&BL Lac&0.007&BL Lac&0.0058&BL Lac&0.0013\\
J0031.5-5648&J003135.1-564640&BL Lac&0.008&Ambiguous&0.0102&-&-\\
J0057.9+6326&J005758.1+632642&BL Lac&0.006&BL Lac&0.0042&BL Lac&0.0001\\
J0156.3-2420&J015624.6-242003&BL Lac&0.007&Ambiguous&0.0251&-&-\\
J0159.0+3313&J015905.0+331255&BL Lac&0.008&BL Lac&0.0008&-&-\\
J0231.0+3505&J023112.2+350445&BL Lac&0.006&BL Lac&0.0022&-&-\\
J0301.6-5617&J030115.1-561644&BL Lac&0.008&Ambiguous&0.011&-&-\\
J0302.5+3354&J030226.7+335448&BL Lac&0.007&BL Lac&0.0077&-&-\\
J0327.6+2620&J032737.2+262008&BL Lac&0.007&BL Lac&0.0072&-&-\\
J0409.2+2542&J040921.6+254440&BL Lac&0.008&BL Lac&0.0014&-&-\\
J0610.8-4911&J061031.8-491222&BL Lac&0.007&BL Lac&0.0002&-&-\\
J0610.8-4911&J061100.0-491034&BL Lac&0.007&BL Lac&0.0098&-&-\\
J0620.7-5034&J062045.7-503350&BL Lac&0.007&BL Lac&0.0019&BL Lac&0.0\\
J0633.9+5840&J063400.1+584036&BL Lac&0.009&BL Lac&0.0028&BL Lac&0.0005\\
J0650.6+2055&J065035.4+205557&BL Lac&0.006&BL Lac&0.0002&BL Lac&0.0003\\
J0704.3-4829&J070421.8-482648&BL Lac&0.009&BL Lac&0.0037&-&-\\
J0800.9+0733&J080056.5+073235&BL Lac&0.007&BL Lac&0.0005&BL Lac&0.0035\\
J0827.0-4111&J082705.4-411159&BL Lac&0.009&BL Lac&0.0016&-&-\\
J0838.5+4013&J083903.0+401546&BL Lac&0.006&BL Lac&0.0005&BL Lac&0.0\\
J0903.5+4057&J090342.8+405503&BL Lac&0.009&Ambiguous&0.0128&Ambiguous&0.0606\\
J0910.1-1816&J091003.9-181613&BL Lac&0.007&Ambiguous&0.0113&BL Lac&0.0002\\
J0914.5+6845&J091429.7+684509&BL Lac&0.007&BL Lac&0.0023&BL Lac&0.0007\\
J0928.4-5256&J092818.7-525701&BL Lac&0.007&BL Lac&0.0001&-&-\\
J0930.9-3030&J093057.9-303118&BL Lac&0.007&BL Lac&0.0085&-&-\\
J1011.1-4420&J101132.0-442255&BL Lac&0.007&BL Lac&0.0007&-&-\\
J1016.1-4247&J101620.7-424723&BL Lac&0.007&BL Lac&0.0006&-&-\\
J1024.5-4543&J102432.5-454428&BL Lac&0.007&BL Lac&0.0005&BL Lac&0.0005\\
J1048.4-5030&J104824.2-502941&BL Lac&0.006&BL Lac&0.0008&-&-\\
J1146.0-0638&J114600.8-063851&BL Lac&0.006&BL Lac&0.0009&BL Lac&0.0006\\
J1155.2-1111&J115514.7-111125&BL Lac&0.007&Ambiguous&0.0104&BL Lac&0.0048\\
J1220.1-2458&J122014.5-245949&BL Lac&0.007&BL Lac&0.0007&BL Lac&0.0002\\
J1243.7+1727&J124351.8+172645&BL Lac&0.007&BL Lac&0.0023&Ambiguous&0.0127\\
J1545.0-6642&J154458.9-664147&BL Lac&0.006&BL Lac&0.0003&BL Lac&0.0\\
J1557.2+3822&J155711.9+382032&BL Lac&0.008&BL Lac&0.0013&-&-\\
J1631.8+4144&J163146.7+414633&BL Lac&0.006&BL Lac&0.0005&BL Lac&0.0033\\
J1651.7-7241&J165151.5-724310&BL Lac&0.007&BL Lac&0.001&-&-\\
J1720.6-5144&J172032.7-514413&BL Lac&0.008&BL Lac&0.002&-&-\\
J1910.8+2856&J191052.2+285624&BL Lac&0.007&BL Lac&0.0021&BL Lac&0.0\\
J1910.8+2856&J191059.4+285635&BL Lac&0.006&BL Lac&0.0008&-&-\\
J1918.0+0331&J191803.6+033030&BL Lac&0.006&BL Lac&0.0033&BL Lac&0.0\\
J1927.5+0154&J192731.3+015357&BL Lac&0.006&Ambiguous&0.0316&-&-\\
J2046.9-5409&J204700.7-541245&BL Lac&0.007&BL Lac&0.009&-&-\\
J2109.6+3954&J210936.4+395513&BL Lac&0.007&BL Lac&0.0013&-&-\\
J2114.9-3326&J211452.1-332533&BL Lac&0.007&BL Lac&0.0005&-&-\\
J2159.6-4620&J215936.1-461953&BL Lac&0.006&BL Lac&0.0025&BL Lac&0.0017\\
J2207.1+2222&J220704.1+222232&BL Lac&0.007&Ambiguous&0.0644&BL Lac&0.006\\
J2225.8-0804&J222552.9-080416&BL Lac&0.008&Ambiguous&0.0269&-&-\\
J2237.2-6726&J223709.3-672614&BL Lac&0.008&BL Lac&0.0023&-&-\\
J2303.9+5554&J230351.7+555618&BL Lac&0.006&BL Lac&0.0003&-&-\\
J2317.7+2839&J231740.0+283954&BL Lac&0.007&BL Lac&0.0005&BL Lac&0.0022\\
&&&&&&&\\
\hline
\end{tabular}
\label{tab:classification-Res1}
\end{center}
\end{table}
\begin{table}
\begin{center}
\caption{Table \ref{tab:classification-Res1} continues...}
\begin{tabular}{rccccccc}
\hline
	&&&&&&\\
	    \multicolumn{2}{c}{Source Name}  & \multicolumn{2}{c}{K23}        &\multicolumn{2}{c}{Present work}&\multicolumn{2}{c}{Present work}\\
	    \multicolumn{2}{c}{}  & \multicolumn{2}{c}{}        &\multicolumn{2}{c}{XGB Model}&\multicolumn{2}{c}{XGBz model}\\
	&&&&&&&\\
\hline
	&&&&&&&\\
	4FGL Name & UVOT Name & Assigned class  &FSRQness  & Assigned class& FSRQness& Assigned Class& FSRQness\\
	&&&&&&\\
\hline
	&&&&&&\\
J0004.4-4001&J000434.2-400035&Ambiguous&0.018&BL Lac&0.0022&BL Lac&0.0002\\
J0025.4-4838&J002536.9-483810&Ambiguous&0.014&BL Lac&0.0081&-&-\\
J0037.2-2653&J003729.5-265045&Ambiguous&0.597&Ambiguous&0.9722&-&-\\
J0058.3-4603&J005806.3-460419&Ambiguous&0.011&BL Lac&0.0041&-&-\\
J0118.3-6008&J011824.0-600753&Ambiguous&0.035&BL Lac&0.0023&-&-\\
J0120.2-7944&J011914.7-794510&Ambiguous&0.947&FSRQ&0.993&-&-\\
J0125.9-6303&J012548.1-630245&Ambiguous&0.022&BL Lac&0.0099&-&-\\
J0209.8+2626&J020946.5+262528&Ambiguous&0.101&BL Lac&0.0031&Ambiguous&0.0114\\
J0240.2-0248&J024004.6-024505&Ambiguous&0.406&Ambiguous&0.5483&-&-\\
J0259.0+0552&J025857.6+055244&Ambiguous&0.01&BL Lac&0.0028&-&-\\
J0406.2+0639&J040607.7+063919&Ambiguous&0.244&Ambiguous&0.0351&-&-\\
J0427.8-6704&J042749.6-670435&Ambiguous&0.141&Ambiguous&0.0194&-&-\\
J0537.5+0959&J053745.9+095759&Ambiguous&0.938&Ambiguous&0.3998&-&-\\
J0539.2-6333&J054002.9-633216&Ambiguous&0.015&Ambiguous&0.0691&-&-\\
J0544.8+5209&J054424.5+521513&Ambiguous&0.959&Ambiguous&0.9793&-&-\\
J0738.6+1311&J073843.4+131330&Ambiguous&0.695&Ambiguous&0.0194&-&-\\
J0800.1-5531&J075949.3-553254&Ambiguous&0.52&Ambiguous&0.9582&-&-\\
J0800.1-5531&J080013.1-553408&Ambiguous&0.29&Ambiguous&0.0839&-&-\\
J0906.1-1011&J090616.1-101426&Ambiguous&0.015&Ambiguous&0.5557&-&-\\
J0934.5+7223&J093333.7+722101&Ambiguous&0.3&Ambiguous&0.3194&-&-\\
J0938.8+5155&J093834.8+515453&Ambiguous&0.014&Ambiguous&0.8298&Ambiguous&0.9516\\
J1008.2-1000&J100848.6-095450&Ambiguous&0.972&Ambiguous&0.0117&Ambiguous&0.0195\\
J1016.2-5729&J101625.7-572807&Ambiguous&0.461&Ambiguous&0.0864&-&-\\
J1018.1-2705&J101750.2-270550&Ambiguous&0.647&Ambiguous&0.0444&-&-\\
J1018.1-4051&J101801.4-405519&Ambiguous&0.951&FSRQ&0.9983&-&-\\
J1018.1-4051&J101807.6-404408&Ambiguous&0.966&Ambiguous&0.9846&-&-\\
J1034.7-4645&J103438.7-464405&Ambiguous&0.017&Ambiguous&0.0387&BL Lac&0.0002\\
J1049.8+2741&J104938.8+274213&Ambiguous&0.012&BL Lac&0.0006&BL Lac&0.0004\\
J1106.7+3623&J110636.5+362650&Ambiguous&0.963&FSRQ&0.9998&FSRQ&0.9991\\
J1111.4+0137&J111114.2+013431&Ambiguous&0.125&BL Lac&0.0076&BL Lac&0.0024\\
J1119.9-1007&J111948.4-100707&Ambiguous&0.029&Ambiguous&0.154&-&-\\
J1122.0-0231&J112213.7-022914&Ambiguous&0.191&BL Lac&0.0027&-&-\\
J1256.8+5329&J125630.5+533205&Ambiguous&0.984&FSRQ&0.9914&FSRQ&0.9914\\
J1320.3-6410&J132015.9-641349&Ambiguous&0.71&Ambiguous&0.0646&-&-\\
J1326.0+3507&J132622.2+350625&Ambiguous&0.055&Ambiguous&0.4274&BL Lac&0.0071\\
J1326.0+3507&J132544.4+350450&Ambiguous&0.028&BL Lac&0.0061&-&-\\
J1415.9-1504&J141546.1-150229&Ambiguous&0.015&BL Lac&0.0046&-&-\\
J1429.8-0739&J142949.5-073305&Ambiguous&0.36&Ambiguous&0.0879&-&-\\
J1513.0-3118&J151244.8-311647&Ambiguous&0.496&BL Lac&0.0029&-&-\\
J1514.8+4448&J151451.0+444957&Ambiguous&0.831&Ambiguous&0.802&-&-\\
J1528.4+2004&J152835.8+200421&Ambiguous&0.013&BL Lac&0.0013&BL Lac&0.0012\\
J1623.7-2315&J162334.1-231750&Ambiguous&0.958&Ambiguous&0.9703&-&-\\
J1644.8+1850&J164457.2+185150&Ambiguous&0.015&Ambiguous&0.2714&-&-\\
J1645.0+1654&J164459.8+165513&Ambiguous&0.031&BL Lac&0.0031&-&-\\
J1650.9-4420&J165124.2-442142&Ambiguous&0.692&Ambiguous&0.1157&-&-\\
J1818.5+2533&J181831.2+253707&Ambiguous&0.721&Ambiguous&0.0104&-&-\\
J1846.9-0227&J184650.7-022904&Ambiguous&0.469&BL Lac&0.0011&-&-\\
J1955.3-5032&J195512.5-503012&Ambiguous&0.487&Ambiguous&0.3922&-&-\\
J2008.4+1619&J200827.6+161844&Ambiguous&0.014&Ambiguous&0.0146&-&-\\
J2041.1-6138&J204112.0-613949&Ambiguous&0.012&BL Lac&0.0078&-&-\\
J2222.9+1507&J222253.9+151055&Ambiguous&0.013&Ambiguous&0.0163&-&-\\
J2240.3-5241&J224017.7-524113&Ambiguous&0.013&BL Lac&0.0005&BL Lac&0.0006\\
J2247.7-5857&J224745.0-585501&Ambiguous&0.552&Ambiguous&0.7538&-&-\\
J2311.6-4427&J231145.6-443221&Ambiguous&0.073&Ambiguous&0.0531&-&-\\
J2326.9-4130&J232653.1-412711&Ambiguous&0.938&Ambiguous&0.6716&-&-\\
J2336.9-8427&J233627.1-842648&Ambiguous&0.01&BL Lac&0.0041&-&-\\
J2337.7-2903&J233730.2-290241&Ambiguous&0.017&BL Lac&0.0075&-&-\\
J2351.4-2818&J235136.5-282154&Ambiguous&0.159&BL Lac&0.0023&-&-\\	
\hline
\end{tabular}
\label{tab:classification-Res2}
\end{center}
\end{table}

\begin{table}
\begin{center}
\caption{List of sources that change their classification after using redshift as an input parameter.}
\begin{tabular}{rccccccc}
\hline
	&&&&&&\\
	    \multicolumn{2}{c}{Source Name}  &\multicolumn{2}{c}{XGB Model} &\multicolumn{2}{c}{XGBz Model}& (Difference in FSRQness)$\times$100\\
	&&&&&&\\
\hline
	&&&&&&\\
	4FGL Name & UVOT Name & Assigned class& FSRQness& Assigned Class& FSRQness\\
	&&&&&&\\
\hline

	&&&&&&\\
	J1637.5+3005&J163728.1+300953&Ambiguous&0.9468&FSRQ&0.9972	&5.04\\
	J0910.1-1816&J091003.9-181613&Ambiguous&0.0113&BL Lac&0.0002	&0.93\\
	J1155.2-1111&J115514.7-111125&Ambiguous&0.0104&BL Lac&0.0048	&0.56\\
	J2207.1+2222&J220704.1+222232&Ambiguous&0.0644&BL Lac&0.006	&5.84\\
	J1034.7-4645&J103438.7-464405&Ambiguous&0.0387&BL Lac&0.0002	&3.85\\
	J1326.0+3507&J132622.2+350625&Ambiguous&0.4274&BL Lac&0.0071	&42.03\\
	\hline
	J1637.5+3005&J163739.2+301013&FSRQ&0.9951&Ambiguous&0.9747	&2.04\\
	J0209.8+2626&J020946.5+262528&BL Lac&0.0031&Ambiguous&0.0114	&1.13\\
	J1243.7+1727&J124351.8+172645&BL Lac&0.0023&Ambiguous&0.0127	&1.04\\
	&&&&&&\\
\hline
\end{tabular}
\label{tab:z_Compare}
\end{center}
\end{table}

\bibliographystyle{mnras}
\bibliography{MS} 
\bsp	
\label{lastpage}
\end{document}